\documentclass[12pt]{article}
\usepackage{graphicx}
\begin{document}
\bigskip
\centerline{\Large \bf Simulation of Galam's contrarian opinions }

\medskip
\centerline{\Large \bf on percolative lattices}

\bigskip
D. Stauffer$^1$ and J.S. S\'a Martins$^2$

\bigskip
Laboratoire PMMH, Ecole Sup\'erieure de Physique et Chimie
Industrielle, 10 rue Vauquelin, F-75231 Paris, Euroland

\medskip

$^1$ Visiting from Institute for Theoretical Physics, Cologne
University, D-50923 K\"oln, Euroland;
stauffer@thp.uni-koeln.de

$^2$ Visiting from Instituto de F\'{\i}sica, Universidade
Federal Fluminense; Av. Litor\^{a}nea s/n, Boa Viagem,
Niter\'{o}i 24210-340, RJ, Brazil; jssm@if.uff.br
\bigskip

Abstract: Galam's model of people voting always against the majority is shown 
to give for the quenched case different results than the original annealed 
model. For people diffusing on a lattice, Galam's phase transitions happen 
only at higher concentrations of people.

\bigskip

Galam \cite{g1} suggested a model of opinion formation with a small minority of
what he called ``contrarians'', who always have the opinion opposite of
that of the majority. He found a phase transition without contrarians such
that the opinion which is initially shared by more than half of the agents
after sufficiently many iterations is shared by everyone: Total polarization.
With a small fraction $a$ of contrarians the phase transition stays at
1/2 but leads no longer to total polarization. Above a critical fraction
$a_c = 1/6$ the phase transition vanishes, and the two possible opinions
are shared each by half the agents. (Since $a$ and $1-a$ seem to be equivalent
we take $a \le 1/2.$) Analogously, spinodal decomposition
in a Glauber-kinetic Ising model gives a spontaneous magnetization of 100
percent at zero temperaure $T$, a spontaneous magnetization below 100 \% for
$0 < T < T_c$, and no spontaneous magnetization for $T$ above the critical
temperature $T_c$. We now test if this analogy between contrarians and
temperature remains valid if we get rid of Galam's restriction that always
three people meet and form a majority opinion. Our generalization then also
allows to study the effect of a bias \cite{g2}, that one of the two opinions
(``status quo'') is taken in the case of a tie vote. The simulation methods
are similar to \cite{stauffer}.

\begin{figure}[hbt]
\begin{center}
\includegraphics[angle=-90,scale=0.5]{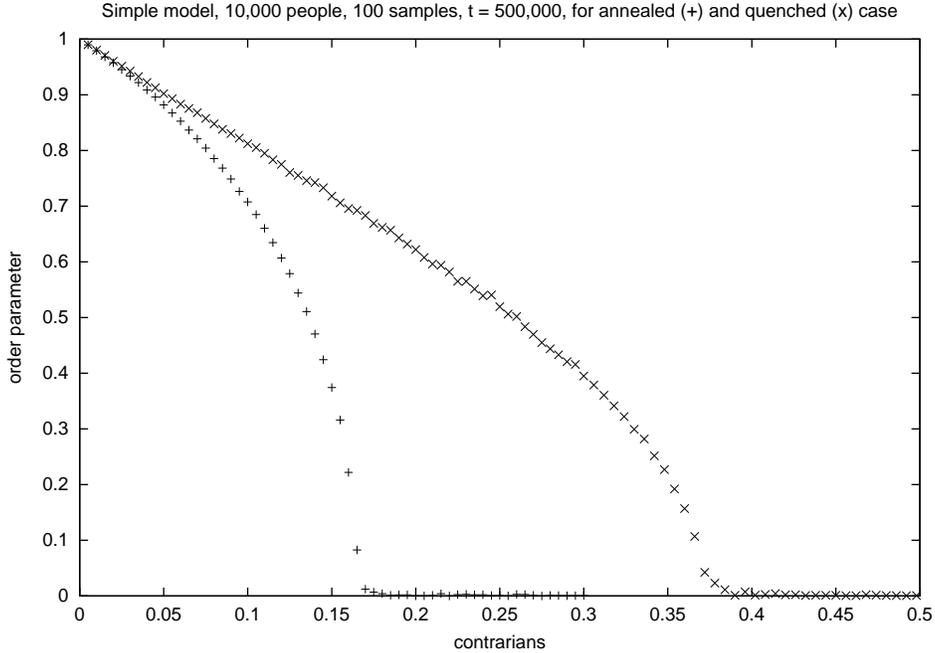}
\end{center}
\caption{
Comparison of annealed and quenched versions of simple Galam model. The order 
parameter $M$ is the number of individuals with +1 opinion minus that of --1 
opinion, divided by the total population.
}
\end{figure}

\begin{figure}[hbt]
\begin{center}
\includegraphics[angle=-90,scale=0.5]{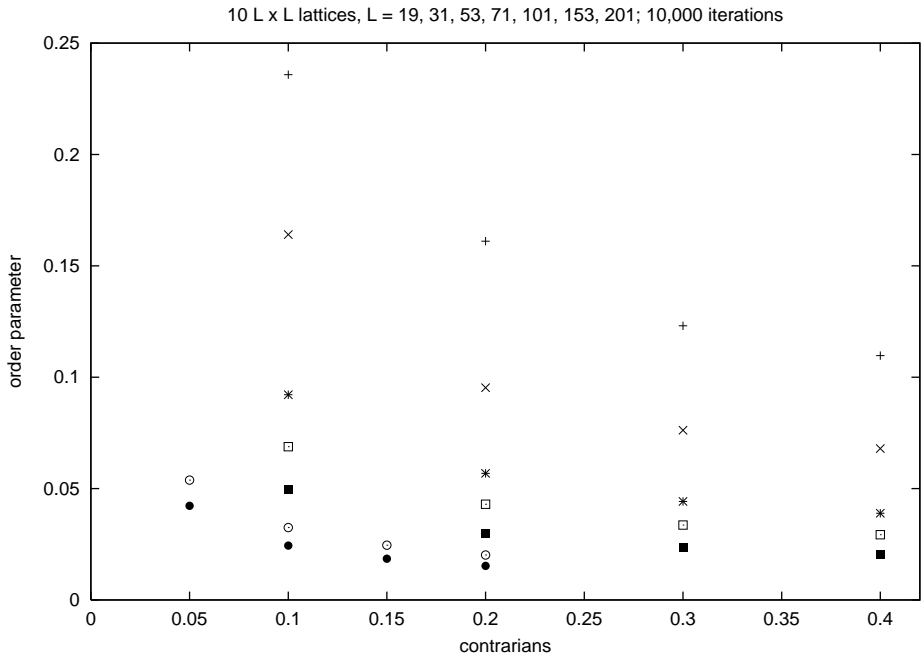}
\end{center}
\caption{
Order parameter $M$ (normalized root-mean-square magnetization) versus
fraction $a$ of contrarians on square lattices of various sizes (see
headline) at concentration $1/4$.
}
\end{figure}

\begin{figure}[hbt]
\begin{center}
\includegraphics[angle=-90,scale=0.5]{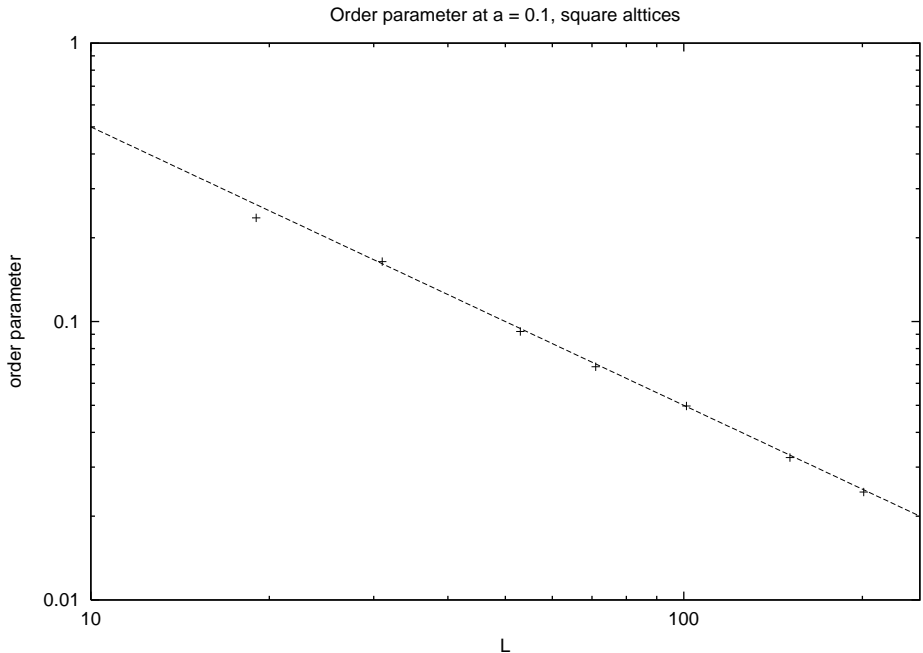}
\end{center}
\caption{
Log-log plot of $M$ versus $L$ at $a=0.1$ (data from Fig.1); the straight
line gives $M = 5/L$.
}
\end{figure}

\begin{figure}[hbt]
\begin{center}
\includegraphics[angle=-90,scale=0.5]{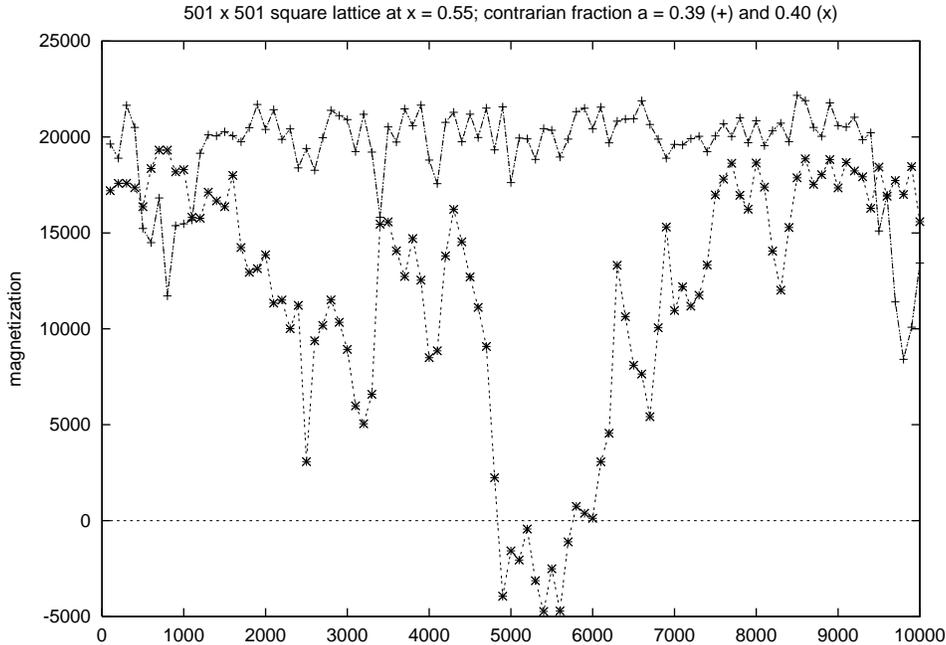}
\end{center}
\caption{
Time variation of the unnormalized magnetization (difference in the number of
opinion + minus opinion $-$) for one sample of $501 \times 501$ at $x=0.55$
for contrarian fractions $0.39$ (+) and $0.40$ (x), thus giving $a_c \sim 0.40$.
With one or ten instead of three diffusion steps per voting step the results do 
not change much.
}
\end{figure}

\begin{figure}[hbt]
\begin{center}
\includegraphics[angle=-90,scale=0.5]{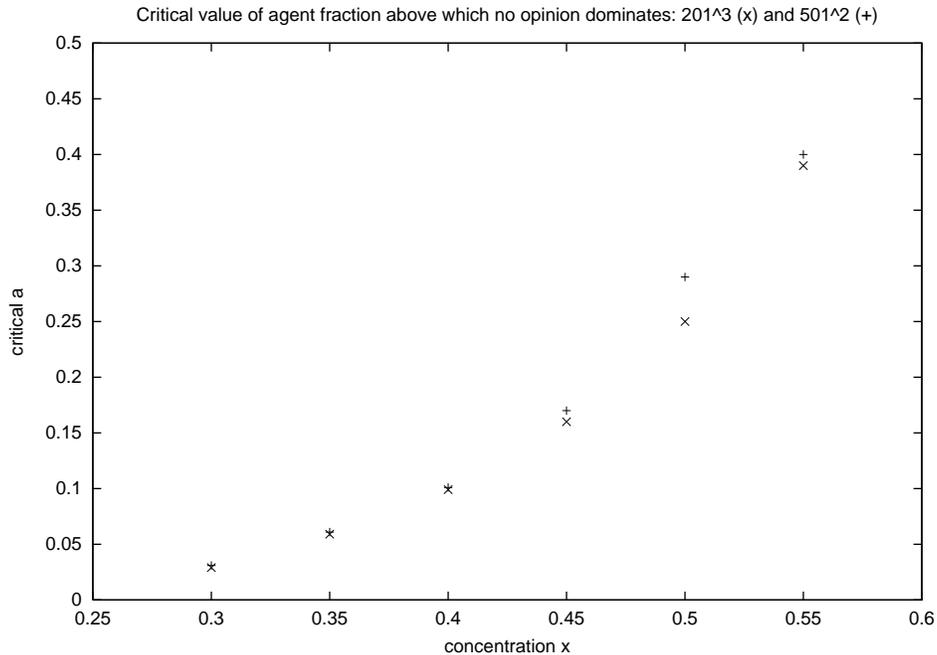}
\end{center}
\caption{
Critical contrarian fraction $a_c$ above which no opinion dominates, versus
agent concentration $x$ on $501 \times 501$ (+) and $201 \times 201$ (x) square
lattices.
}
\end{figure}

\begin{figure}[hbt]
\begin{center}
\includegraphics[angle=-90,scale=0.5]{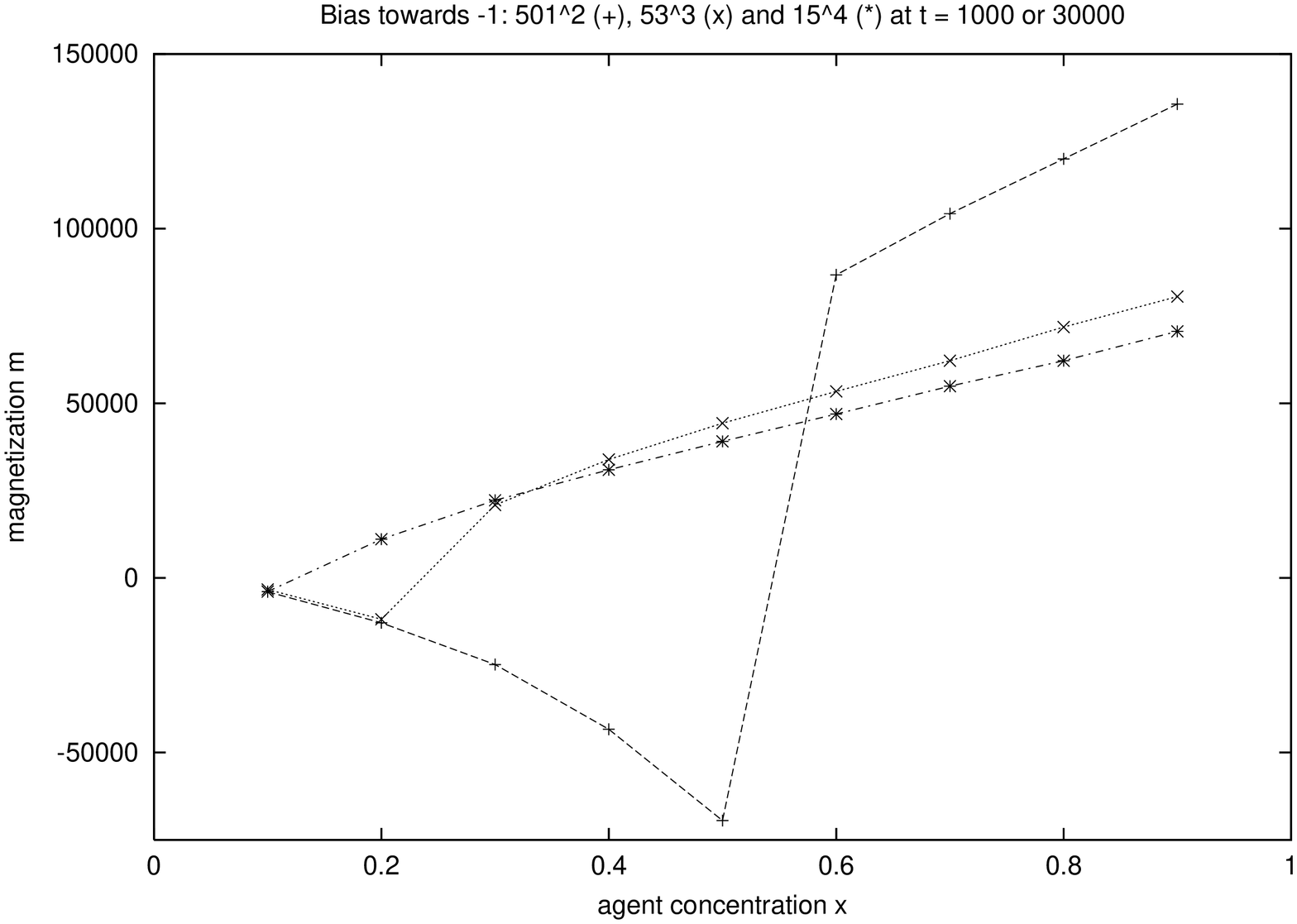}
\end{center}
\caption{
Unnormalized magnetization $m$ after $1000$ iterations with bias, versus 
agent concentration $x$, at fixed contrarian fraction $a = 1/5$ in two, 
three and four dimensions
}
\end{figure}

First, we present a Monte Carlo simulation of the Galam contrarian model, 
for two different implementations of contrarianism. At the start, every agent 
has one of two possible opinions (spins) $\pm 1$ 
selected randomly, with a given initial probability of, say, the up opinion. At 
each time step, $3$ agents are selected at random and form a local group; their 
opinions are then updated accordingly to a majority rule: winner takes all, and 
all $3$ agents have the same opinion at the end. 
Contrarians are those agents that disagree with the majority, and after an 
update end up with the minority opinion. In the annealed implementation of 
contrarians, for which Ref. \cite{g1} presents an analytical solution, 
when a local group is to have its opinion updated, a fixed fraction $a$ of 
the agents is randomly chosen to be contrarian. An individual that is tossed 
as a contrarian in one time step may act as a non-contrarian in the next. 
For a more realistic quenched contrarianism version, on the other hand, a 
fraction $a$ of the population is randomly selected as contrarians at 
the beginning of the simulation, and they keep this character throughout the 
dynamics.

The results, shown in Figure 1, confirm the analytically derived critical 
fraction of contrarians if these are annealed ($a_c = 0.17$). Above this 
concentration, no majority is formed in the long run and one gets the 
hung scenario refered to in \cite{g1}. For the quenched version, the 
hung scenario is still present, but now the critical fraction of contrarians 
is much higher. Our estimate for its value is $a_c = 0.39$.

All runs had the same parameters: the total population was $10,000$, results 
were averaged over the final $10,000$ steps out of a total of $500,000$ updates,
and also averaged over $100$ different samples, with an initial probability 
of the $+1$ opinion (up spin) equal to $0.45$ as in \cite{g1}.

After this simulation close to the original Galam model, we now present Monte 
Carlo simulations of interacting agents diffusing on a square lattice. 
Initially we distribute agents randomly with concentration $x$ on a hypercubic
lattice in $d$ dimensions. Thus they form small percolative
clusters with $s = 1, 2, 3, \dots$ sites. (In Galam's analytically solved
model, all clusters had the same odd number of agents, like $s=3$.) Every
agent has one of two possible opinions $\pm 1$, selected randomly, just like 
in the original model. At each iteration every agent makes
(typically) three diffusion steps, i.e. it selects randomly one of the $2d$
nearest neighbours and moves there if the site is empty. The moving agent
carries its opinion with it. In this way a new
percolative cluster distribution is formed. Then each cluster of agents
determines the majority opinion within that cluster; and all agents within
that cluster are then convinced by this opinion. In the case of a tie vote
for even $s$, impossible for the $s=3$ case of \cite{g1} but possible in
\cite{g2,stauffer}, the majority opinion is found randomly if we simulate no
bias, and it is $-1$ or ``status quo'' if we simulate bias. After this exchange
of opinions for all clusters, one iteration is finished and the next starts with
diffusion of agents, as above.

The contrarian people are stubborn agents, forming a random fraction $a$ of
all agents, and carrying their stubbornness with them while diffusing: 
quenched. 
Their opinion is always opposite to that of the cluster to which they belong,
even if that opinion was determined randomly. Similar effects were simulated
in the Sznajd model \cite{Sznajd}.

Our simulations without contrarians and without bias gave in all ten $L \times
L$ square lattices a
complete consensus, either all opinions $+1$ or all opinions $-1$, about
equally often; the average time to reach this consensus increases from 225 for
$L=11$ to about 27,000 for $L = 71$. With a small fraction of contrarians
on small lattices still a complete consensus is reached, but this artifact
vanishes for a larger number of contrarians and a finite observation time.
The ``magnetization'' $m$ is the number of up opinions minus the number of
down opinions and fluctuates wildly, having positive and negative values.
We thus take its root-mean-square $M$ as order parameter, $M^2  = <m^2>/L^d$
where the brackets indicate a time average. We also average $M$ in Fig. 2 
over all time intervals and samples. Fig. 3 repeats some of these data to show
that $M$ vanishes as $1/L$, as is expected from fluctuations in finite
two-dimensional systems. Thus for $x = 1/4$ this percolation model does
not show a positive order parameter like a spontaneous magnetization, for small
$a$, and thus also no upper critical fraction of contrarians, in contrast
to the simple model of \cite{g1}.

An entirely different picture evolves if the concentration $x$ of agents on the
square lattice is doubled from $x=1/4$ to $x=1/2$. Now, starting from half
the opinions up and half down, after thousands of iterations a wide consensus
evolves: one opinion is much more frequent that the other; thus we have
ferromagnetism in the Ising-model analogy, for not too large temperature (=
contrarian fraction $a$). Then it is more practical to start with
90 percent of the opinions up and 10 percent down and to check whether this
initial majority survives (ferromagnetism) or is replaced by about equal
numbers of $+1$ and $-1$. Fig. 4 shows an example of how we determined the
critical $a_c$, and Fig. 5 summarizes the resulting phase diagram $a_c(x)$:
Similar to a dilute Ising model, we have ferromagnetism (widespread consensus)
only above some minimum agent concentration $x \simeq 0.3$ and then only for low
enough contrarian fractions $a$. (This method of starting with positive order 
parameters and watching them changing their signs works if the system size 
goes to infinity for a fixed long observation time.)

For $x$ above the percolation threshold of $0.593$ we find ferromagnetism 
with small fluctuations in the magnetization, except near $a = 1/2$ (not shown); 
apparently the initial orientation of the infinite cluster dominates the 
system.

For three dimensions, at $x = 1/10$ the critical contrarian fraction is about 
$0.03$ while at $x = 1/6$ it is near $0.09$; in four dimensions the corresponding 
$a_c$ values are $0.09$ and $0.26$.

Very different are the results when the bias is switched on, such that a cluster
with evenly divided opinions always agrees on the status quo $(-1)$. Starting again 
with $90 \%$ of the votes at $+1$, we arrive at a majority for $-1$ at low 
concentrations $x$, while for $x$ above the percolation threshold the majority 
stays at $+1$, as shown in Fig. 6.

In summary, only for large enough agent concentrations $x$, but below the 
percolation threshold, do our results look similar to those of Galam \cite{g1}. 
The infinite cluster for $x$ above the percolation threshold prevents the 
emergence of another opinion and might be interpreted as a dictatorship.

\medskip
\noindent {\bf Acknowledgements}: We thank  PMMH at ESPCI for the warm
hospitality and Sorin T\u{a}nase-Nicola for help with the computer facilities. 
JSSM acknowledges funding from the Brazilian agencies CNPq and FAPERJ.


\newpage

\begin{thebibliography}{99}
\bibitem{g1}  S. Galam, eprint cond-mat/0307404

\bibitem{g2}  S. Galam, Eur. Phys. J. B 25, 403 (2002)

\bibitem{stauffer} D. Stauffer, Int. J. Mod. Phys.  C 13, 975 (2002)

\bibitem{Sznajd} Sznajd-Weron K, and Sznajd, J., Int. J. Mod. Phys. C 11,
1157 (2000); J. Schneider, poster at Sociophysics conference, Bielefeld,
May 2002.

\end{thebibliography}
\end{document}